\documentclass[letterpaper, 10pt, conference]{ieeeconf}
\IEEEoverridecommandlockouts
\usepackage{graphicx, enumerate}
\usepackage[OT1]{fontenc}
\usepackage{cite,array, bm, svg}
\usepackage{dsfont, epstopdf}
\usepackage{mathrsfs, subfigure, color}
\usepackage{amssymb, multirow}
\usepackage[cmex10]{amsmath}
\usepackage{subeqnarray, tabularx}
\usepackage{cases, stmaryrd}
\usepackage{algpseudocode}
\usepackage[ruled, vlined, linesnumbered]{algorithm2e}
\usepackage[utf8]{inputenc}
\usepackage{threeparttable}
\usepackage{stackengine}
\usepackage{verbatim}
\newcommand\xrowht[2][0]{\addstackgap[.5\dimexpr#2\relax]{\vphantom{#1}}}

\DeclareFontFamily{U}{FdSymbolC}{}
\DeclareFontShape{U}{FdSymbolC}{m}{n}{<-> s * FdSymbolC-Book}{}
\DeclareSymbolFont{fdarrows}{U}{FdSymbolC}{m}{n}
\DeclareMathSymbol{\vDdash}{\mathrel}{fdarrows}{254}

\DeclareFontFamily{U}{FdSymbolD}{}
\DeclareFontShape{U}{FdSymbolD}{m}{n}{<-> s * FdSymbolD-Book}{}
\DeclareSymbolFont{fdnarrows}{U}{FdSymbolD}{m}{n}
\DeclareMathSymbol{\nvDdash}{\mathrel}{fdnarrows}{254}
\makeatother

\definecolor{bluencs}{rgb}{0.0, 0.53, 0.74}

\newtheorem{theorem}{Theorem}

\newtheorem{proposition}{Proposition}
\newtheorem{definition}{Definition}

\DeclareMathOperator{\ess}{\mathrm{ess}}

\DeclareMathOperator{\cstr}{\mathrm{c}}

\DeclareMathOperator{\enab}{\mathrm{enab}}

\DeclareMathOperator{\reach}{\textbf{Reach}}

\begin{document}

\title{\LARGE \bf Zonotope-based Controller Synthesis for LTL Specifications
\thanks{This work was supported by the European Research Council (ERC) Grant 864017--L2C, the CHIST-ERA 2018 project DRUID-NET, the Walloon Region and the Innoviris Foundation. J. Calbert is an FRIA fellow, and R. Jungers is an FNRS Research Associate.}
}

\author{Wei~Ren, Julien Calbert and Rapha\"el Jungers
\thanks{W. Ren, J. Calbert and R. Jungers are with ICTEAM Institute, Universit\'{e} catholique de Louvain, 1348 Louvain-la-Neuve, Belgium. Email: \texttt{\{w.ren, julien.calbert, raphael.jungers\}@uclouvain.be}.}
}

\maketitle

\begin{abstract}
This paper studies the controller synthesis problem for Linear Temporal Logic (LTL) specifications using (constrained) zonotope techniques. First, we implement (constrained) zonotope techniques to partition the state space and further to verify whether the LTL specification can be satisfied. Once the LTL specification can be satisfied, the next step is to design a controller to guarantee the satisfaction of the LTL specification for dynamic systems. Based on the verification of the LTL specification, an abstraction-based control design approach is proposed in this paper: a novel abstraction construction is developed first, then finite local abstract controllers are designed to achieve the LTL specification, and finally the designed abstract controllers are combined and refined as the controller for the original system. The proposed control strategy is illustrated via a numerical example from autonomous robots.
\end{abstract}

%%%%%%%%%%%%%%%%%%%%%%%%%%%%%%%%%%%%%%%%%%%%%%%%%%%%%%%%%%%%%%%%%%%%%%%%%%%%%%%%%%%%%%%%%%%%%%%%%%%%%%%%%%%%%%%%%%%%%%%%%%%%%%%%%%%%%%%%%%%%%%%%%%%%%%%%%%%%%%%%%%%%
\section{Introduction}
\label{sec-intro}
%%%%%%%%%%%%%%%%%%%%%%%%%%%%%%%%%%%%%%%%%%%%%%%%%%%%%%%%%%%%%%%%%%%%%%%%%%%%%%%%%%%%%%%%%%%%%%%%%%%%%%%%%%%%%%%%%%%%%%%%%%%%%%%%%%%%%%%%%%%%%%%%%%%%%%%%%%%%%%%%%%%%

In terms of high-level specifications like linear temporal logic (LTL) formulas \cite{Baier2008principles}, the control synthesis problem is generally not easy to be solved directly on continuous dynamics. One approach is based on discrete abstractions \cite{Milner1989communication, Tabuada2006linear}, which allow to deal with controller synthesis problems efficiently via techniques developed in the fields of supervisory control \cite{Ramadge1987supervisory} or game theory \cite{Ramadge1987modular}. In this approach, a symbolic abstraction is first constructed for the continuous system such that certain behavioral relationship between them is satisfied, and then formal methods from the field of computer science are leveraged to synthesize a controller for the symbolic abstraction to satisfy the high-level specifications. This bottom-up control approach has been applied to study different systems and specifications \cite{Girard2010approximately, Zamani2012symbolic, Pola2008approximately}.

Once the behavioral relation is guaranteed, the controller for the symbolic abstraction can be refined as the controller for the continuous dynamics \cite{Pola2008approximately, Girard2012controller, Ren2020symbolic, Reissig2017feedback}. Hence, the basis and essence of the abstraction-based control approach is the abstraction construction. Recently, many construction methods have been proposed. In terms of behavioral relations, (bi)simulation relation and its variants were applied in \cite{Pola2008approximately, Girard2010approximately, Girard2007approximation}, and feedback refinement relation was proposed in \cite{Reissig2017feedback}. In terms of the space partition, different partition approaches were proposed, like quantization techniques \cite{Pola2008approximately, Ren2019dynamic, Reissig2017feedback, Ren2020symbolic}, interval techniques \cite{Li2017invariance}, and reachability techniques \cite{Meyer2017compositional}. These methods are based on the approximation of the whole state space, which may result in huge complexity, and constrain system dynamics and the desired specifications; see e.g., \cite{Reissig2017feedback, Meyer2017compositional} for more details.

To deal with these limitations, we propose a novel control synthesis approach for nonlinear dynamic systems under LTL specifications. The proposed approach is based on the top-down design methodology \cite{Russell2020artificial}, and thus is opposed to the bottom-up control approach in previous works. The approach consists of two steps: the specification verification and the controller design. The first step facilitates the second step by dividing the state space and allowing finite granular abstractions, whereas the second step leverages the first step in order to derive the corresponding controller. First, following the top-down design methodology and based on zonotope techniques, a novel approach is developed for the partition of the state space and the verification of the LTL specification. To be specific, the state space is divided via finite (constrained) zonotopes, which are allowed to intersect with each other, and the intersection relation among these (constrained) zonotopes results in an undirected graph. Given the initial state space, the satisfactability of the LTL specification can be verified based on the resulting graph. The specification verification leads to an ordered sequence of chosen (constrained) zonotopes to be applied in the controller synthesis, and a decomposition of the global specification into finite local specifications. Different from the existing partition approaches \cite{Fainekos2009temporal, Ren2019dynamic} where all cells are disjoint, our approach allows different cells to be intersected, and this intersection relation plays an essential role in both the specification verification and the controller design.

The second step is to explore the controller for the LTL specification based on the abstraction-based control approach. To show this, we consider a chosen (constrained) zonotope. A local symbolic abstraction is first constructed for the current (constrained) zonotope such that certain equivalence relation between the original system and the constructed abstraction holds. Then, the intersection region between the current and previous (constrained) zonotopes is the initial region and the intersection region between the current and next (constrained) zonotopes is the target region. Finally, the local controller is designed via the local abstraction. In particular, the initial state set is the initial region for the first chosen (constrained) zonotope. Here, the applied approximation technique is based on the properties of zonotopes, which is different from the quantization technique \cite{Girard2007approximation, Ren2020symbolic}. With the equivalence relation between the original system and the constructed abstraction, all designed local controllers are combined and refined as the controller for the original system with the LTL specification.

Preliminaries and problem formulation are introduced in Section \ref{sec-notepre}. The zonotope-based partition is proposed in Section \ref{sec-partition} and LTL specifications are verified in Section \ref{sec-verifydiscrete}. The abstraction-based control design approach is proposed in Section \ref{sec-abstractionbased}. A numerical example is given in Section \ref{sec-example}. Conclusion and future work are presented in Section \ref{sec-conclusion}.

%%%%%%%%%%%%%%%%%%%%%%%%%%%%%%%%%%%%%%%%%%%%%%%%%%%%%%%%%%%%%%%%%%%%%%%%%%%%%%%%%%%%%%%%%%%%%%%%%%%%%%%%%%%%%%%%%%%%%%%%%%%%%%%%%%%%%%%%%%%%%%%%%%%%%%%%%%%%%%%%%%%%
\section{Notation and Preliminaries}
\label{sec-notepre}
%%%%%%%%%%%%%%%%%%%%%%%%%%%%%%%%%%%%%%%%%%%%%%%%%%%%%%%%%%%%%%%%%%%%%%%%%%%%%%%%%%%%%%%%%%%%%%%%%%%%%%%%%%%%%%%%%%%%%%%%%%%%%%%%%%%%%%%%%%%%%%%%%%%%%%%%%%%%%%%%%%%%

$\mathbb{R}:=(-\infty, +\infty)$; $\mathbb{R}^{+}_{0}:=[0, +\infty)$; $\mathbb{R}^{+}:=(0, +\infty)$; $\mathbb{N}:=\{0, 1, \ldots\}$; $\mathbb{N}^{+}:=\{1, 2, \ldots\}$. Given $A, B\subset\mathbb{R}^{n}$, $B\backslash A:=\{x: x\in B, x\notin A\}$. $A\subset\mathbb{R}^{n}$ is a connected region if it cannot be represented as the union of two or more disjoint non-empty open regions. Given a vector $x\in\mathbb{R}^{n}$, $x_{i}$ is the $i$-th element of $x$, $|x|$ is the Euclidean norm of $x$, and $\|x\|$ is the infinity norm of $x$. The closed ball centered at $x\in\mathbb{R}^{n}$ with radius $\varepsilon\in\mathbb{R}^{+}$ is defined by $\mathbf{B}(x, \varepsilon)=\{y\in\mathbb{R}^{n}:\|x-y\|\leq\varepsilon\}$. Given a set $\Lambda\subset\mathbb{R}^{n}$, $\Lambda^{\circ}$ is the interior of $\Lambda$; the $\varepsilon$-expansion of $\Lambda$ is $\mathbf{E}_{\varepsilon}(\Lambda):=\{y\in\mathbb{R}^{n}: \exists x\in\Lambda, \|y-x\|\leq\varepsilon\}$. Given a measurable function $f: \mathbb{R}^{+}_{0}\rightarrow\mathbb{R}^{n}$, the (essential) supremum of $f$ is $\|f\|:=\ess\sup\{\|f(t)\|:t\in\mathbb{R}^{+}_{0}\}$. Given $A, B\subset\mathbb{R}^{n}$, a relation $\mathcal{F}\subset A\times B$ is the map $\mathcal{F}: A\rightarrow2^{B}$ defined by $b\in\mathcal{F}(a)$ if and only if $(a, b)\in\mathcal{F}$.

A set $Z\subset\mathbb{R}^{n}$ is a \textit{zonotope}, if there exists $(\mathbf{c}, \mathbf{G})\in\mathbb{R}^{n}\times\mathbb{R}^{n\times n_{g}}$ such that $Z=\{\mathbf{c}+\mathbf{G}\xi: \|\xi\|\leq1\}$. A set $Z^{\cstr}\subset\mathbb{R}^{n}$ is a \textit{constrained zonotope}, if there exists $(\mathbf{c}, \mathbf{G}, \mathbf{A}, \mathbf{b})\in\mathbb{R}^{n}\times\mathbb{R}^{n\times n_{g}}\times\mathbb{R}^{n_{c}\times n_{g}}\times\mathbb{R}^{n_{c}}$ such that $Z^{\cstr}=\{\mathbf{c}+\mathbf{G}\xi: \|\xi\|\leq1, \mathbf{A}\xi=\mathbf{b}\}$. $\mathbf{c}$ is the center, $\mathbf{G}$ is the generator matrix with each column being a generator, and $\mathbf{A}\xi=\mathbf{b}$ is the constraint. From \cite{Scott2016constrained}, $Z^{\cstr}$ is a constrained zonotope if and only if it is a convex polytope, and a convex polytope is a zonotope if and only if every 2-face is centrally symmetric.

\subsection{Transition Systems}
\label{subsec-approbisimu}

\begin{definition}[\cite{Girard2007approximation}]
\label{def-2}
A \textit{transition system} is a quadruple $T = (X, X^{0}, U, \Delta)$ with: (i) a state set $X$; (ii) a set of initial states $X^{0}\subseteq X$; (iii) a input set $U$; (iv) a transition relation $\Delta\subseteq X\times U\times X$. $T$ is \textit{symbolic} if $X$ and $U$ are countable.
\end{definition}

The transition $(x, u, x')\in\Delta$ is denoted by $x'\in\Delta(x, u)$, which means that the system can evolve from the state $x$ to the state $x'$ under the input $u$. An input $u\in U$ belongs to \textit{the set of the enabled inputs} at the state $x$, denoted by $\enab(x)$, if $\Delta(x, u)\neq\varnothing$. $T$ is \textit{deterministic} if for all $x\in X$ and all $u\in\enab(x)$, $\Delta(x, u)$ has exactly one element. In this case, we write $x'=\Delta(x, u)$ with a slight abuse of notation.

\begin{definition}[\cite{Reissig2017feedback}]
\label{def-3}
Let $T_{i}=(X_{i}, X^{0}_{i}, U_{i}, \Delta_{i})$ with $i\in\{1, 2\}$ be two transition systems, and $U_{2}\subseteq U_{1}$. A relation $\mathcal{F}\subseteq X_{1}\times X_{2}$ is a \textit{feedback refinement relation} from $T_{1}$ to $T_{2}$, if for all $(x_{1}, x_{2})\in\mathcal{F}$, (a) $U_{2}(x_{2})\subseteq U_{1}(x_{1})$; (b) $u\in U_{2}(x_{2})\Rightarrow\mathcal{F}(\Delta_{1}(x_{1}, u))\subseteq\Delta_{2}(x_{2}, u)$, where $U_{i}(x):=\{u\in U_{i}: u\in\enab(x)\}$. Denote by $T_{1}\preceq_{\mathcal{F}}T_{2}$ if $\mathcal{F}\subseteq X_{1}\times X_{2}$ is a feedback refinement relation from $T_{1}$ to $T_{2}$.
\end{definition}

\begin{definition}[\cite{Pola2008approximately}]
\label{def-4}
A \textit{control system} is a quadruple $\Sigma=(X, U, \mathcal{U}, f)$, where, (i) $X\subset\mathbb{R}^{n}$ is the state set; (ii) $U\subseteq\mathbb{R}^{m}$ is the input set; (iii) $\mathcal{U}$ is a subset of all piecewise continuous functions of time from the interval $[0, \infty)\subset\mathbb{R}$ to $U$; (iv) $f: X\times U\rightarrow\mathbb{R}^{n}$ is a continuous map satisfying the Lipschitz assumption: there exists a constant $L\in\mathbb{R}^{+}$ such that for all $x, y\in X$ and all $u\in U$, $\|f(x, u)-f(y, u)\|\leq L\|x-y\|$. Given any $b>a\geq0$, a locally absolutely continuous curve $\xi: (a, b)\rightarrow\mathbb{R}^{n}$ is a \textit{trajectory} of $\Sigma$, if there exists $\mathbf{u}\in\mathcal{U}$ such that $\dot{\xi}(t)=f(\xi(t), \mathbf{u}(t))$ for almost all $t\in(a, b)$.
\end{definition}

The trajectory defined on $[0, \tau]$ with $\tau\in\mathbb{R}^{+}$ is denoted as $\mathbf{x}: [0, \tau]\rightarrow X$. The trajectory starting from the initial time $t_{0}$ is denoted by $x[t_{0}]$. Denote by $\mathbf{x}(t, x, u)$ the point reached at $t\in(a, b)$ under the input $u$ from $x$. Such a point is determined uniquely from the assumptions on $f$.

\subsection{Linear Temporal Logic Specifications}
\label{subsec-dynquan}

Let $\Pi:=\{\pi_{1}, \ldots, \pi_{\mathfrak{n}}\}$ be a finite set of atomic propositions with $\mathfrak{n}\in\mathbb{N}^{+}$, and $\Phi_{\Pi}$ be the set of all Boolean combinations of elements of $\Pi$. The denotation $\llbracket\cdot\rrbracket: \Pi\rightarrow\mathscr{P}(X)$ of each symbol in $\Pi$ is a subset of $X$, i.e., $\llbracket\pi\rrbracket\subseteq X$ for any $\pi\in\Pi$. Here, $\mathscr{P}(\Gamma)$ denotes the powerset of a set $\Gamma$.

Based on atomic propositions, Boolean connectors like negation $\neg$ and conjunction $\wedge$, and two temporal operators $\bigcirc$ (`next') and $\textsf{U}$ (`until'), Linear Temporal Logic (LTL) is formed via the following syntax \cite{Baier2008principles}: $\varphi::=\top\mid\alpha\mid\neg\varphi\mid\varphi_{1}\wedge\varphi_{2}\mid\bigcirc\varphi\mid\varphi_{1}\textsf{U}\varphi_{2}$, where $\varphi, \varphi_{1}, \varphi_{2}$ are LTL formulas. The Boolean connector disjunction $\vee$, and temporal operators $\lozenge$ (`eventually') and $\square$ (`always') can be derived below: $\varphi_{1}\vee\varphi_{2}:=\neg(\neg\varphi_{1}\wedge\neg\varphi_{2})$, $\lozenge\varphi:=\top\textsf{U}\varphi$ and $\square\phi:=\neg\lozenge\neg\phi$. Formal definitions for the LTL semantics and model checking can be found in \cite{Baier2008principles}. For the system $\Sigma$, if the LTL formula $\phi$ is satisfied over the trajectory $x[t_{0}]$ with respect to the atomic proposition mapping $\llbracket\cdot\rrbracket$, then we denote by $(x[t_{0}], \llbracket\cdot\rrbracket)\models\phi$, otherwise, $(x[t_{0}], \llbracket\cdot\rrbracket)\not\models\phi$. For the system $\Sigma$ and the LTL formula $\phi$, we aim to deal with the following two problems.
\begin{enumerate}[1]
  \item Given the system $\Sigma$ and an LTL formula $\phi$, verify whether the formula $\phi$ could be satisfied.
  \item If the LTL formula $\phi$ is satisfied, design the controller $C$ such that the LTL formula $\phi$ is satisfied for the system $\Sigma$.
\end{enumerate}
By solving the first problem, we verify the satisfaction of the LTL specification in the given workspace in Section \ref{sec-verifydiscrete}. If the verification is feasible, then the controller is designed for the system $\Sigma$ to solve the second problem in Section \ref{sec-abstractionbased}.

%%%%%%%%%%%%%%%%%%%%%%%%%%%%%%%%%%%%%%%%%%%%%%%%%%%%%%%%%%%%%%%%%%%%%%%%%%%%%%%%%%%%%%%%%%%%%%%%%%%%%%%%%%%%%%%%%%%%%%%%%%%%%%%%%%%%%%%%%%%%%%%%%%%%%%%%%%%%%%%%%%%%
\section{Zonotopes based Partition}
\label{sec-partition}
%%%%%%%%%%%%%%%%%%%%%%%%%%%%%%%%%%%%%%%%%%%%%%%%%%%%%%%%%%%%%%%%%%%%%%%%%%%%%%%%%%%%%%%%%%%%%%%%%%%%%%%%%%%%%%%%%%%%%%%%%%%%%%%%%%%%%%%%%%%%%%%%%%%%%%%%%%%%%%%%%%%%

To solve the above two problems, we first propose a novel method for the state-space partition. The partition strategy is presented in Algorithm \ref{alg-1}. Line 1 is to generate finite zonotopes and constrained zonotopes to cover the state space. Since the generated zonotopes and constrained zonotopes are not necessarily overlapped, Line 2 is to expand all generated zonotopes and constrained zonotopes to ensure each (constrained) zonotope to overlap with its neighbour zonotopes and constrained zonotopes. In Line 2, the expansion operator is applied to all generated zonotopes and constrained zonotopes. We emphasize that Line 2 plays important roles in the controller synthesis, which will be explained in Section \ref{sec-abstractionbased} in detail.

\begin{algorithm}[!t]
\DontPrintSemicolon
\small
\caption{State Space Partition}
\label{alg-1}
\KwIn{the state space $X\subset\mathbb{R}^{n}$, the precision $\varepsilon>0$}
\KwOut{the set of zonotopes $\mathbf{Z}$}
Generate finite zonotopes $Z_{i}$ and constrained zonotopes $Z^{\cstr}_{j}$ to cover the state space $X$\;
Expand both $Z_{i}$ and $Z^{\cstr}_{j}$ via the precision $\varepsilon>0$\;
\textbf{return} $\mathbf{Z}=(\cup^{N}_{i=1}\mathbf{E}_{\varepsilon}(Z_{i}))\cup(\cup^{M}_{j=1}\mathbf{E}_{\varepsilon}(Z^{\cstr}_{j}))$
\end{algorithm}

\begin{figure}[!t]
\begin{center}
\begin{picture}(80, 90)
\put(-75, -12){\resizebox{80mm}{40mm}{\includegraphics[width=2.5in]{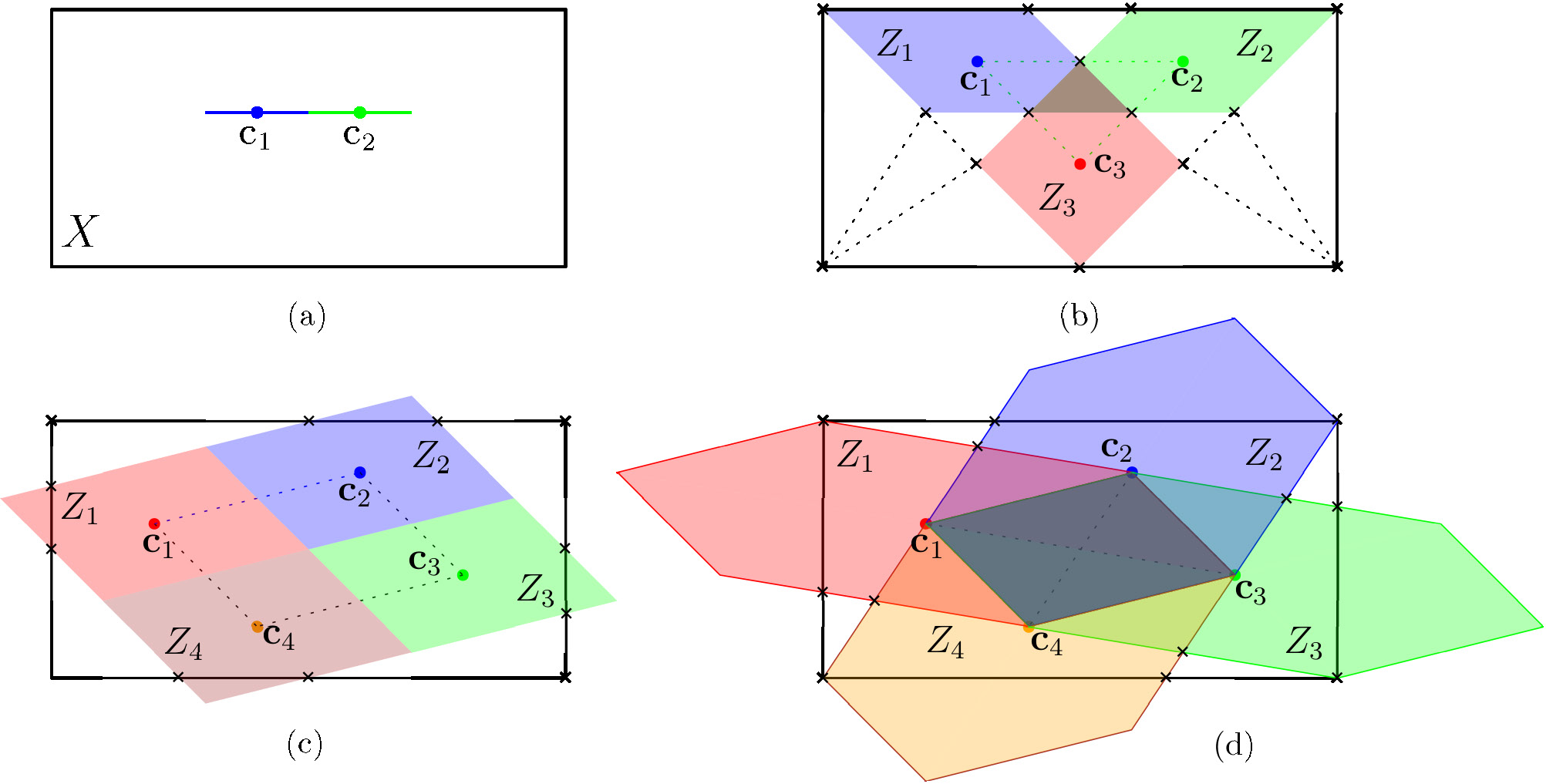}}}
\end{picture}
\end{center}
\caption{Illustration of the zonotope generation. (a) $N=2$: the generated zonotopes are 2 segments and thus not well-defined. (b)$N=3$: 3 zonotopes are generated and overlapped. (c) $N=4$ and each center connects with 2 neighbor centers: 4 zonotopes are not overlapped. (d) $N=4$ and each center connects with 3 neighbor centers: 4 zonotopes are overlapped.}
\label{fig-1}
\end{figure}

%\subsection{Generation of Zonotopes and Constrained Zonotopes}
%\label{subsec-generatezonotope}

In the following, we establish Line 1 of Algorithm \ref{alg-1}, which is formulated in Algorithm \ref{alg-2}. To begin with, the number $N\in\mathbb{N}$ of zonotopes to be generated is set \textit{a priori}, and then we choose $N\in\mathbb{N}$ points $\mathbf{c}_{i}\in X$ arbitrarily as the centers of zonotopes, where $i\in\mathcal{N}_{1}:=\{1, \ldots, N\}$. Here, we assume that $N\in\mathbb{N}$ satisfies $N>n$ with $n\in\mathbb{N}^{+}$ being the dimension of the state space. These points are connected such that each point is connected with at least $n\in\mathbb{N}$ neighbour points. That is, these connections lead to at least $n\in\mathbb{N}$ vectors for each point, which is used as the generators for each zonotope. With these centers and generators, we can generate $N$ zonotopes as in Line 3 in Algorithm \ref{alg-2}.

In the zonotope generation, the constrains on the choice of the centers (i.e., $N>n$ and $\mathbf{G}_{i}$ is full-rank) are to guarantee the construction of zonotopes. To show this, we consider a two-dimensional space $X\subset\mathbb{R}^{2}$ as in Fig. \ref{fig-1}. If we choose $N=2$ as in Fig. \ref{fig-1}(a), then the generated zonotopes are two segments. If $\mathbf{G}_{i}$ is not full-rank, then each center connects with only one neighbour center or some centers are in the same line, and the generated zonotopes are segments either. Therefore, the lower bound of $N\in\mathbb{N}$ and the full-rank condition on $\mathbf{G}_{i}$ are to ensure the well-definedness of the zonotope generation. Note that different $N\in\mathbb{N}$ and $\mathbf{G}_{i}$ have effects on the generated zonotopes; see Fig. \ref{fig-1}.

\begin{algorithm}[!t]
\DontPrintSemicolon
\small
\caption{Zonotope Generation}
\label{alg-2}
\KwIn{the state domain $X\subset\mathbb{R}^{n}$, the integer $N>n$}
\KwOut{finite zonotopes and constrained zonotopes}
Choose $N$ points in $X$ randomly as the centers $\mathbf{c}_{i}\in\mathbb{R}^{n}$\;
Connect these points such that for each $\mathbf{c}_{i}\in\mathbb{R}^{n}$, there exists a full-rank matrix:
\begin{equation}
\label{eqn-1}
\mathbf{G}_{i}=(\mathbf{c}_{k_{1}}-\mathbf{c}_{i}, \ldots, \mathbf{c}_{k_{i}}-\mathbf{c}_{i}), \quad k_{i}\geq n
\end{equation}\;
\vspace*{-\baselineskip}
With $\mathbf{c}_{i}\in\mathbb{R}^{n}$ and $0.5\mathbf{G}_{i}$, construct the zonotope
\begin{equation}
\label{eqn-2}
Z_{i}=\{\mathbf{c}_{i}+0.5\mathbf{G}_{i}\xi: \|\xi\|\leq1\}
\end{equation}\;
\vspace*{-\baselineskip}
\eIf{$X\setminus(\cup^{N}_{i=1}Z_{i})=\varnothing$}{
No need to construct constrained zonotopes
}{
Determine the set $\mathbb{V}_{1}$ of intersection among $\partial Z_{i}, \ldots, \partial Z_{N}, \partial X$\;
Refine $\mathbb{V}_{1}$ as $\mathbb{V}$ by excluding those in $\cup^{N}_{i=1}Z^{\circ}_{i}$\;
Generate the set $\mathbb{S}_{1}$ of regions by connecting vertices in $\mathbb{V}$\;
Refine $\mathbb{S}_{1}$ as $\mathbb{S}$ by excluding those intersecting with $\cup^{N}_{i=1}Z_{i}$\;
Construct $M$ constrained zonotopes $\cup^{M}_{j=1}Z^{\cstr}_{j}\supseteq\mathbb{S}$\;
}
\textbf{return} $\mathbf{Z}=(\cup^{N}_{i=1}Z_{i})\cup(\cup^{M}_{j=1}Z^{\cstr}_{j})$
\end{algorithm}

If the union of all generated zonotopes covers the state space, then there is no need for the generation of constrained zonotopes; otherwise, constrained zonotopes are needed. The motivation of using constrained zonotopes is that constrained zonotopes are allowed to be asymmetric \cite{Scott2016constrained, Rego2020guaranteed} and thus can be used to cover asymmetric regions. The construction rule is based on intersection points among the generated zonotopes and the state space, and presented explicitly as follows: we first determine the intersection vertices among boundaries of the generated zonotopes and the state space, then rule out those in the interiors of at least one of the generated zonotopes and achieve the vertices to generate constrained zonotopes (see the crosses in Fig. \ref{fig-1}), and finally connect these vertices to generate regions that do not intersect with the generated zonotopes. These regions are represented as constrained zonotopes.

The generation of constrained zonotopes relies on basic operations of zonotopes. For instance, the vertices of the generated zonotopes can be obtained via the transformation of zonotopes into the V-representation  \cite{Kochdumper2019representation}. These vertices are connected to determine the intersection vertices, which form the set $\mathbb{V}_{1}$ in Line 7 of Algorithm \ref{alg-2}. The set $\mathbb{V}_{1}$ is refined as the set $\mathbb{V}$ in Line 8 by excluding those in the generated zonotopes. Based on the vertices in $\mathbb{V}$ and using the transformation from V-representation into Z-representation \cite{Kochdumper2019representation}, constrained zonotopes are generated in Lines 9-11 of Algorithm \ref{alg-2} to cover the region $X\setminus(\cup^{N}_{i=1}Z_{i})$.

All generated zonotopes and constrained zonotopes are labeled via a symbol set $\Pi:=\{\pi_{1}, \ldots, \pi_{N}, \pi_{N+1}, \ldots,$ $\pi_{N+M}\}$, where $\llbracket\pi_{i}\rrbracket=\mathbf{E}_{\varepsilon}(Z_{i})$ for all $i\in\mathcal{N}_{1}$, $\llbracket\pi_{N+j}\rrbracket=\mathbf{E}_{\varepsilon}(Z^{\cstr}_{j})$ for all $j\in\mathcal{N}_{2}:=\{1, \ldots, M\}$. Let $\mathcal{N}:=\{1, \ldots, N+M\}$, and the partition of the state space $X$ is $\mathbf{Z}:=\{\mathbf{Z}_{k}: \mathbf{Z}_{k}=\llbracket\pi_{k}\rrbracket, k\in\mathcal{N}\}$.

%%%%%%%%%%%%%%%%%%%%%%%%%%%%%%%%%%%%%%%%%%%%%%%%%%%%%%%%%%%%%%%%%%%%%%%%%%%%%%%%%%%%%%%%%%%%%%%%%%%%%%%%%%%%%%%%%%%%%%%%%%%%%%%%%%%%%%%%%%%%%%%%%%%%%%%%%%%%%%%%%%%%
\section{Verification of LTL Formulas}
\label{sec-verifydiscrete}
%%%%%%%%%%%%%%%%%%%%%%%%%%%%%%%%%%%%%%%%%%%%%%%%%%%%%%%%%%%%%%%%%%%%%%%%%%%%%%%%%%%%%%%%%%%%%%%%%%%%%%%%%%%%%%%%%%%%%%%%%%%%%%%%%%%%%%%%%%%%%%%%%%%%%%%%%%%%%%%%%%%%

With the partition of the state space, the LTL formula is verified in this section. For this purpose, we denote by $\mathcal{O}:=\{\mathcal{O}_{k}: k\in\mathbb{J}\}\subset X$ all forbidden regions including obstacles and states that are not allowed to be visited, where $\mathbb{J}\subset\mathbb{N}$ is finite. A region $A\subset X$ is \textit{admissible} if $A\setminus\mathcal{O}$ is a connected region. First, we verify whether all intersection regions are admissible and derive all admissible intersection regions, which further imply the relation among zonotopes and constrained zonotopes. Second, this relation can be transformed into an undirected graph via the adjacency matrix. This two-step mechanism is summarized as Algorithm \ref{alg-3}.

\begin{algorithm}[!t]
\DontPrintSemicolon
\small
\caption{Adjacency Matrix}
\label{alg-3}
\KwIn{the partition $\mathbf{Z}$, the forbidden region $\mathcal{O}\subset X$}
\KwOut{the matrix $\Upsilon=[a_{ij}]$ and the set $\mathbf{I}=\{\mathbf{I}_{ij}\}$}
\For{$i=1:1:N+M-1$}{
\eIf{$\mathbf{Z}_{i}\cap\mathcal{O}\neq\mathbf{Z}_{i}$}{
\For{$j=1:1:N+M$}{
\eIf{$\mathbf{Z}_{j}\cap\mathcal{O}\neq\mathbf{Z}_{j}$}{
$\Omega=\mathbf{Z}_{i}\cap\mathbf{Z}_{j}$\;
\eIf{$\Omega=\varnothing$}{
$a_{ij}=0$\;}{
\eIf{$(\mathbf{Z}_{i}\cup\mathbf{Z}_{j})\setminus(\Omega\cap\mathcal{O})$ is connected}{
$a_{ij}=1  \quad   \mathbf{I}_{ij}=\Omega\setminus\mathcal{O}$\;
}{
$a_{ij}=0$\;
}
}
}{$a_{ij}=0$\;}
}
}{$a_{ij}=0$ for $j\in\mathcal{N}$\;}
}
\textbf{return} $\Upsilon=[a_{ij}], \mathbf{I}=\{\mathbf{I}_{ij}\}$
\end{algorithm}

\begin{theorem}
\label{thm-1}
Algorithm \ref{alg-3} terminates in finite time. The obtained adjacency matrix $\Upsilon$ shows all admissible intersections among all zonotopes and constrained zonotopes.
\end{theorem}

From the adjacency matrix $\Upsilon$, we generate a graph $\mathcal{G}=(\mathcal{V}, \mathcal{E})$, where the vertex set $\mathcal{V}=\Pi$, and the edge set is $\mathcal{E}\subseteq\Pi\times\Pi$ with $(\pi_{i}, \pi_{j})\in\mathcal{E}$ if $a_{ij}=1$. To include the initial state region and the LTL formula $\phi$, the graph $\mathcal{G}$ is generalized. To be specific, the initial state region is contained in the constrained zonotope $Z^{\cstr}_{0}$ with the symbol $\pi_{0}$. From the LTL formula $\phi$, the regions of interest are obtained and denoted as finite constrained zonotopes $Z^{\cstr}_{\phi,\ell}$ with the symbol $\pi_{\phi,\ell}$, where $\ell\in\mathbb{K}:=\{1, \ldots, \mathrm{K}\}$ and $\mathrm{K}\in\mathbb{N}$. Both $Z^{\cstr}_{0}$ and $Z^{\cstr}_{\phi,\ell}$ may intersect with some elements of $\mathbf{Z}$, which can be verified easily. Hence, $\mathcal{G}$ is generalized as $\bar{\mathcal{G}}=(\bar{\mathcal{V}}, \bar{\mathcal{E}})$ with the vertex set $\bar{\mathcal{V}}=\Pi\cup\{\pi_{0}, \pi_{\phi,1}, \ldots, \pi_{\phi,\mathrm{K}}\}$ and the edge set $\bar{\mathcal{E}}\subseteq\bar{\mathcal{V}}\times\bar{\mathcal{V}}$ extending $\mathcal{E}$ by adding the connections among $Z^{\cstr}_{0}, Z^{\cstr}_{\phi,\ell}$ and $\mathbf{Z}$. On the other hand, we can use any standard LTL model checker \cite{Baier2008principles} to solve the LTL planning problem and to derive an accepting path $\bar{\pi}:=\{\bar{\pi}_{0}, \bar{\pi}_{1}, \bar{\pi}_{2}, \ldots\}$ with $\bar{\pi}_{0}=\pi_{0}$ and $\bar{\pi}_{k}$ with $k\in\mathbb{N}^{+}$ from $\{\pi_{\phi,1}, \ldots, \pi_{\phi,\mathrm{K}}\}$.

In the graph $\bar{\mathcal{G}}$, each path $\mathbf{P}$ can be projected into a sequence of finite (constrained) zonotopes denoted as $\mathbf{Z}_{\textsf{p}}:=\{\llbracket\pi\rrbracket\in\mathbf{Z}: \pi\in\mathbf{P}\}$. In addition, the intersection region set in $\mathbf{Z}_{\textsf{p}}$ is denoted as $\mathbf{I}_{\textsf{p}}:=\{\mathbf{I}^{\textsf{p}}_{ij}=\mathbf{Z}_{\textsf{p}i}\cap\mathbf{Z}_{\textsf{p}j}\in\mathbf{I}: \forall \mathbf{Z}_{\textsf{p}i}, \mathbf{Z}_{\textsf{p}j}\in\mathbf{Z}_{\textsf{p}}\}$. The following theorem is derived to verify the satisfaction of the LTL formula.

\begin{theorem}
\label{thm-2}
Consider the state space $X\subseteq\mathbb{R}^{n}$, the initial state region $X_{0}\subset Z^{\cstr}_{0}$, and the LTL formula $\phi$. The following two statements are equivalent.
\begin{enumerate}
  \item The LTL formula $\phi$ can be satisfied in $X\subseteq\mathbb{R}^{n}$.
  \item In $\bar{\mathcal{G}}$ there exists a path $\mathbf{P}$ realizing the accepting path $\bar{\pi}$. For each $\mathbf{Z}_{\textsf{p}i}\in\mathbf{Z}_{\textsf{p}}$ with $\mathbf{Z}_{\textsf{p}}$ from the path $\mathbf{P}$, either of the following conditions holds:
  \begin{enumerate}
    \item $\mathbf{Z}_{\textsf{p}i}\setminus{\mathcal{O}}$ is a connected region;

    \item otherwise, there exists a connected subregion $\bar{\mathbf{Z}}_{\textsf{p}i}\subset\mathbf{Z}_{\textsf{p}i}\setminus\mathcal{O}$ such that $\bar{\mathbf{Z}}_{\textsf{p}i}\cap\mathbf{I}^{\textsf{p}}_{(i-1)i}\cap\mathbf{I}^{\textsf{p}}_{i(i+1)}\neq\varnothing$, and $\bar{\mathbf{Z}}_{\textsf{p}i}\cap(\cup_{\ell\in\mathbb{K}}Z^{\cstr}_{\ell})\neq\varnothing$ if $\mathbf{Z}_{\textsf{p}i}\cap(\cup_{\ell\in\mathbb{K}}Z^{\cstr}_{\ell})\neq\varnothing$.
\end{enumerate}
\end{enumerate}
\end{theorem}

In terms of the state space, Theorem \ref{thm-2} shows how to verify the LTL formula $\phi$ via the graph theory, and thus solves Problem 1. Note that the path $\mathbf{P}$ is not necessarily unique.

%%%%%%%%%%%%%%%%%%%%%%%%%%%%%%%%%%%%%%%%%%%%%%%%%%%%%%%%%%%%%%%%%%%%%%%%%%%%%%%%%%%%%%%%%%%%%%%%%%%%%%%%%%%%%%%%%%%%%%%%%%%%%%%%%%%%%%%%%%%%%%%%%%%%%%%%%%%%%%%%%%%%%%
\section{Abstraction-based Controller Synthesis}
\label{sec-abstractionbased}
%%%%%%%%%%%%%%%%%%%%%%%%%%%%%%%%%%%%%%%%%%%%%%%%%%%%%%%%%%%%%%%%%%%%%%%%%%%%%%%%%%%%%%%%%%%%%%%%%%%%%%%%%%%%%%%%%%%%%%%%%%%%%%%%%%%%%%%%%%%%%%%%%%%%%%%%%%%%%%%%%%%%%%

To solve Problem 2, the abstraction-based control techniques are applied, and the whole design process consists of four steps: the time discretization of the original system is first derived; the state and input spaces are approximated; the symbolic abstraction is constructed; and finally the abstract controller is designed and refined into the controller for the original system for the LTL specification.

The time-discretization of the system $\Sigma$ is established via the sampling technique. Let the sampling period be $\tau>0$ as a design parameter, and the sampled-data system is written as a transition system $T_{\tau}(\Sigma):=(X_{1}, X^{0}_{1}, U_{1}, \Delta_{1})$, where,
\begin{itemize}
\item the set of states is $X_{1}=X$;
\item the set of initial states is $X^{0}_{1}=X$;
\item the set of inputs is $U_{1}=\{u\in\mathcal{U}: \mathbf{x}(\tau, x, u) \text{ is defined }$ $\text{for all } x\in X\}$;
\item the transition relation is as follows: for $x\in X_{1}$ and $u\in U_{1}$, $x'=\Delta_{1}(x, u)$ if and only if $x'=\mathbf{x}(\tau, x, u)$.
\end{itemize}
The system $T_{\tau}(\Sigma)$ is deterministic. Note that $T_{\tau}(\Sigma)$ can be treated as the time approximation of the system $\Sigma$.

\subsection{Approximation of State and Input Spaces}
\label{subsec-approximate}

With the time-discretization and the partition of the state domain, the next is to approximate state and input spaces. Without loss of generality, we focus on the approximation of any zonotope $Z\subset X_{1}$ and constrained zonotope $Z^{\cstr}\subset X_{1}$.

\subsubsection{Approximation of Zonotopes} Different from the quantization-based approximation technique in \cite{Girard2007approximation}, the applied approximation technique is based on the properties of zonotopes. Let $Z=\{\mathbf{c}+\mathbf{G}\xi:\|\xi\|\leq1\}\in\mathbf{Z}$ be a zonotope.

First, from the generator $\mathbf{G}$, let $\mathbf{G}_{\textsf{b}}:=(g_{1}, \ldots, g_{\mathrm{L}})=(\mathbf{g}_{1}/N_{1}, \ldots, \mathbf{g}_{\mathrm{L}}/N_{\mathrm{L}})$ be the basic generator, which plays the similar role of the state space discretization parameter as in \cite{Girard2007approximation}. The choice of $\{N_{l}: l\in\mathbb{L}=\{1, \ldots, \mathrm{L}\}\}$ determines the approximation of the state space, and is determined via the desired approximation precision. For instance, given the desired approximation precision $\varepsilon>0$, $\max_{l\in\mathbb{L}}|g_{l}|\leq\varepsilon$ is imposed to constrain the choice of $\{N_{1}, \ldots, N_{\mathrm{L}}\}$. Based on the basic generator $\mathbf{G}_{\textsf{b}}$, we obtain the set $\mathbf{F}:=\cup_{l\in\mathbb{L}}\mathbf{F}_{l}$ with $\mathbf{F}_{l}:=\{\mathbf{c}\pm g_{l}, \ldots, \mathbf{c}\pm\mathbf{g}_{l}, \ldots, \mathbf{c}\pm\bar{N}_{l}g_{l}\}$, where $\bar{N}_{l}\geq N_{l}$ is the largest integer such that $\mathbf{c}\pm\bar{N}_{l}\mathrm{g}_{l}\in\mathbf{Z}$.

Second, based on the set $\mathbf{F}$, the zonotope $Z$ is approximated as follows. Given $p_{lj}\in\mathbf{F}_{l}$, we generate other points in the following way: $p_{lj}\in\mathbf{F}_{l}$ is the basis; any vector in $\{\mathbf{g}_{1}, \ldots,$ $\mathbf{g}_{l-1}, \mathbf{g}_{l+1}, \ldots, \mathbf{g}_{\mathrm{L}}\}$ is the direction; and finally
\begin{align}
\label{eqn-3}
\bar{\mathbf{F}}_{l}:=\bigcup_{0<j\leq\bar{N}_{l}}\bigcup_{k\in\mathbb{L}, k\neq l}\{p_{lj}\pm g_{k}, \ldots, p_{lj}\pm\mathbf{g}_{k}\}\cap Z.
\end{align}
This generation mechanism is terminated until the generated point does not belong to $Z$. That is, $\bar{\mathbf{F}}_{l}\subset Z$. Since $\mathbf{F}_{l}$ is finite, this generation mechanism can be implemented recursively and terminated in finite time. Therefore, $Z$ is approximated by the set $\mathcal{A}(Z):=\{\mathbf{c}\}\cup(\cup_{l\in\mathbb{L}}(\mathbf{F}_{l}\cup\bar{\mathbf{F}}_{l}))$.

\subsubsection{Approximation of Constrained Zonotopes} Following the above approximation for zonotopes, the constrained zonotope $Z^{\cstr}$ can be approximated similarly. The only difference lies in $\mathbf{A}\xi=\mathbf{b}$. Hence, $\bar{\mathbf{F}}_{l}$ in \eqref{eqn-3} is changed to
\begin{align}
\label{eqn-4}
\bar{\mathbf{F}}^{\cstr}_{l}:=\bigcup_{0<j\leq\bar{N}_{l}}\bigcup_{k\in\mathbb{L}, k\neq l}\{p_{lj}\pm g_{k}, \ldots, p_{lj}\pm\mathbf{g}_{k}\}\cap Z^{\cstr}.
\end{align}
Following the same mechanism, we derive $\mathcal{A}(Z^{\cstr})$ to approximate $Z^{\cstr}$. Based on the generator matrix $\mathbf{G}$, we introduce the norm $\|\cdot\|_{\mathbf{G}}$ defined as $\|\mathbf{v}\|_{\mathbf{G}}:=\max_{k\in\mathbb{L}}\{\mathbf{v}\cdot\mathbf{g}_{k}/|\mathbf{g}_{k}|\}$ for $\mathbf{v}\in\mathbb{R}^{n}$, where $\mathbf{v}\cdot\mathbf{g}_{k}/|\mathbf{g}_{k}|$ is the scalar projection of $\mathbf{v}$ onto $\mathbf{g}_{k}$. With the norm $\|\cdot\|_{\mathbf{G}}$, we can see that for any $x\in Z$ (or $x\in Z^{\cstr}$), there exists $q\in\mathcal{A}(Z)$ (or $q\in\mathcal{A}(Z^{\cstr})$) such that $\|x-q\|_{\mathbf{G}}\leq0.5\max_{l\in\mathbb{L}}\{|g_{l}|\}$.

\subsubsection{Approximation of Input Space}
For the zonotope $Z\subset X_{1}$, its input set $U_{1}(Z)\subseteq U_{1}$ is defined as $\cup_{x\in Z}\enab(x)$, and then approximated as follows. Given any $q\in\mathcal{A}(Z)$, the reachable set of $T_{\tau}(\Sigma)$ from $q$ is denoted by $\reach(\tau, q):=\{x'\in Z: \mathbf{x}(\tau, q, u)=x', u\in U_{1}(Z)\}$, which is well defined due to the input set $U_{1}(Z)$. The reachable set $\reach(\tau, q)$ is approximated below. Given any $\eta\in\mathbb{R}^{+}$, consider the set $\mathcal{S}_{\eta}(\tau, q):=\{v\in\mathcal{A}(Z): \exists z\in\reach(\tau, q)$ such that $\|v-z\|_{\mathbf{G}}\leq\eta/2\}$, which is a countable set and where $\eta\in\mathbb{R}^{+}$ is constrained by the approximation precision. Define the function $\psi: \mathcal{S}_{\eta}(\tau, q)\rightarrow U_{1}(Z)$ such that for any $v\in\mathcal{S}_{\eta}(\tau, q)$, there exists an input $u_{1}=\psi(v)\in U_{1}(Z)$ such that $\|v-\mathbf{x}(\tau, q, u_{1})\|_{\mathbf{G}}\leq\eta/2$. We define the set $U_{2}(q):=\psi(\mathcal{S}_{\eta}(\tau, q))$, which captures the set of the inputs applied at the state $q\in\mathcal{A}(Z)$. The set $U_{2}(q)$ is countable since $U_{2}(q)$ is the image of the map $\psi$ of the countable set $\mathcal{S}_{\eta}(\tau, q)$. Therefore, the input set $U_{1}(Z)$ is approximated by mean of the following countable set: $U_{2}(Z):=\cup_{q\in\mathcal{A}(Z)}U_{2}(q)$. That is, given any $q\in\mathcal{A}(Z)$, for any $u_{1}\in U_{1}(Z)$, there exists $u_{2}\in U_{2}(q)$ such that $\|\mathbf{x}(\tau, q, u_{1})-\mathbf{x}(\tau, q, u_{2})\|_{\mathbf{G}}\leq\eta$.

Finally, since all elements of $\mathbf{Z}$ are $\varepsilon$-expansion of (constrained) zonotopes and still are (constrained) zonotopes, the above approximation mechanism can be applied similarly.

\subsection{Abstraction Construction}
\label{subsec-construction}

Now we construct the symbolic abstraction for $T_{\tau}(\Sigma)$ with the state space $Z\in\mathbf{Z}$, which is denoted as $T_{\tau}(\Sigma, Z)$. The symbolic abstraction of $T_{\tau}(\Sigma, Z)$ is described by the transition system $T_{\tau, \eta}(\Sigma, Z)=(X_{2}, X^{0}_{2}, U_{2}, \Delta_{2})$, where,
\begin{itemize}
\item the set of states is $X_{2}=\mathcal{A}(Z)$;
\item the set of initial states is $X^{0}_{2}=\mathcal{A}(Z_{0})$ with $Z_{0}\subseteq Z$;
\item the set of inputs is $U_{2}=U_{2}(Z)$;
\item the transition relation is given by: for $q_{1}, q_{2}\in X_{2}$ and $v\in U_{2}$, $q_{2}\in\Delta_{2}(q_{1}, u)$ if and only if $q_{2}\in\{\bar{q}\in X_{2}: \|\mathbf{x}(\tau, q, v)-\bar{q}\|_{\mathbf{G}}\leq(0.5+e^{L\tau})\varepsilon\}$.
\end{itemize}
From the transition relation, $T_{\tau, \eta}(\Sigma, Z)$ is nondeterministic.

\begin{theorem}
\label{thm-3}
Consider the system $T_{\tau}(\Sigma, Z)$ and its abstraction $T_{\tau, \eta}(\Sigma, Z)$ with the time and input sampling parameters $\tau, \eta\in\mathbb{R}^{+}$. Given a desired precision $\varepsilon\in\mathbb{R}^{+}$, if the map $\mathcal{F}: Z\rightarrow\mathcal{A}(Z)$ is given by $\mathcal{F}(x)=\{q\in\mathcal{A}(Z): \|x-q\|_{\mathbf{G}}\leq\varepsilon\}$,
then $T_{\tau}(\Sigma, Z)\preceq_{\mathcal{F}}T_{\tau, \eta}(\Sigma, Z)$.
\end{theorem}

For each (constrained) zonotope, we can follow the similar mechanism to construct the symbolic abstraction. Hence, the symbolic abstraction is only for the system with the (constrained) zonotope in $\mathbf{Z}$ being the state space, which is different from the construction approach in \cite{Ren2020symbolic, Girard2012controller, Hsu2018multi}.

\subsection{Controller Design and Refinement}
\label{subsec-consyn}

With the path $\mathbf{P}$ from Theorem \ref{thm-2} and the constructed symbolic abstraction, the controller is designed to achieve the LTL specification for the original system. For this purpose, we first denote by $\mathbf{Z}_{\textsf{p}}\subset\mathbf{Z}$ the (constrained) zonotope set from the path $\mathbf{P}$, then have the symbolic abstraction $T_{\tau, \eta}(\Sigma, \mathbf{Z}_{\textsf{p}})$, and finally derive the following proposition.

\begin{proposition}
\label{prop-1}
Given $T_{\tau}(\Sigma, \mathbf{Z}_{\textsf{p}})\preceq_{\mathcal{F}}T_{\tau, \eta}(\Sigma, \mathbf{Z}_{\textsf{p}})$, if there is an abstract controller $C_{\mathrm{a}}: \mathbf{Z}_{\textsf{p}}\rightarrow U_{2}$ for $T_{\tau, \eta}(\Sigma, \mathbf{Z}_{\textsf{p}})$ with the LTL formula $\phi$, then there exists a controller $C(x):=C_{\mathrm{a}}(\mathcal{F}(x))$ for $T_{\tau}(\Sigma, \mathbf{Z}_{\textsf{p}})$ with the LTL formula $\phi$.
\end{proposition}

With Proposition \ref{prop-1}, the abstract controller can be refined as the controller for the system $T_{\tau}(\Sigma)$, and the next is the abstract controller design, which is presented in the following recursive steps. Let $\mathbf{Z}_{\textsf{p}}=\{\mathbf{Z}_{\textsf{p}1}, \ldots, \mathbf{Z}_{\textsf{p}\mathfrak{l}}\}$  with $\mathfrak{l}\in\mathbb{N}^{+}$ be ordered (constrained) zonotopes from the path $\mathbf{P}$.
\begin{enumerate}
  \item For $\mathbf{Z}_{\textsf{p}1}$, the initial set is $(\mathbf{Z}_{\textsf{p}1}\cap Z^{\cstr}_{0})\setminus\mathcal{O}$ and the target set is $(\mathbf{Z}_{\textsf{p}1}\cap\mathbf{Z}_{\textsf{p}2})\setminus\mathcal{O}$.
  \item For $i\in\{2, \ldots, \mathfrak{l}-1\}$, the initial set is $(\mathbf{Z}_{\textsf{p}(i-1)}\cap\mathbf{Z}_{\textsf{p}i})\setminus\mathcal{O}$, and the target set is $(\mathbf{Z}_{\textsf{p}i}\cap\mathbf{Z}_{\textsf{p}(i+1)})\setminus\mathcal{O}$. Note that the final target set is contained in $\mathbf{Z}_{\textsf{p}\mathfrak{l}}$.
  \item Based on the classic fixed-point algorithm \cite{Ramadge1987supervisory}, the abstract controller $C_{i}$ in $\mathbf{Z}_{\textsf{p}i}$ is designed such that the system can move from the initial set to the target set.
  \item All designed abstract controllers $C_{i}$ are combined sequentially as the overall abstract controller $C_{\mathrm{a}}$ for the system $T_{\tau, \eta}(\Sigma, \mathbf{Z}_{\textsf{p}})$ with the LTL specification $\phi$.
\end{enumerate}

%%%%%%%%%%%%%%%%%%%%%%%%%%%%%%%%%%%%%%%%%%%%%%%%%%%%%%%%%%%%%%%%%%%%%%%%%%%%%%%%%%%%%%%%%%%%%
\section{Numerical Example}
\label{sec-example}
%%%%%%%%%%%%%%%%%%%%%%%%%%%%%%%%%%%%%%%%%%%%%%%%%%%%%%%%%%%%%%%%%%%%%%%%%%%%%%%%%%%%%%%%%%%%%

\begin{figure}[!t]
\begin{center}
\begin{picture}(50, 65)
\put(-50, -14){\resizebox{50mm}{25mm}{\includegraphics[width=2.5in]{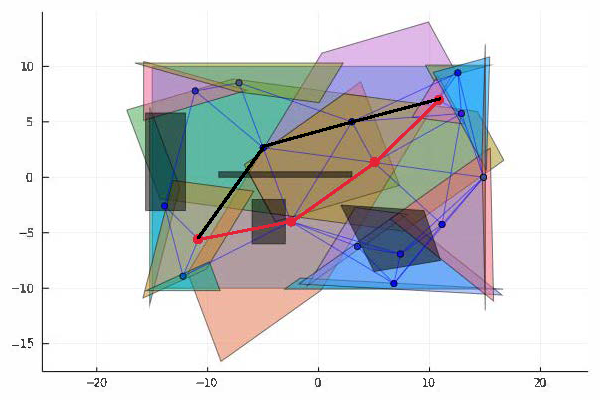}}}
\end{picture}
\end{center}
\caption{\scriptsize{Illustration of the partition of the state space and the graph generation. Both black and red lines are two paths for the LTL formula $\phi$.}}
\label{fig-4}
\end{figure}

\begin{figure}[!t]
\begin{center}
\begin{picture}(50, 50)
\put(-50, -15){\resizebox{50mm}{25mm}{\includegraphics[width=2.5in]{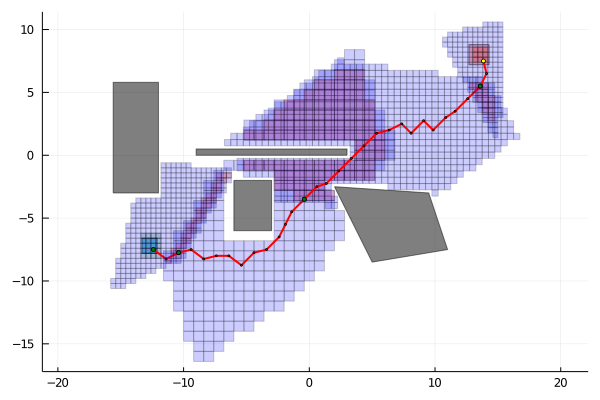}}}
\end{picture}
\end{center}
\caption{\scriptsize{State trajectory based on the abstract controller for the LTL formula $\phi$. The grey regions are obstacles, the dark blue region is the initial state set, the orange region is the target region, the purple regions are intersection regions, and the state trajectory is shown in red line.}}
\label{fig-5}
\end{figure}

The derived results are illustrated in this section. The computation is executed using Julia on a laptop with a 16 GB RAM and an Intel i7-10610U processor (1.80GHz). Consider the autonomous vehicle \cite{Fainekos2009temporal} denoted by $\Sigma: (\dot{x}_{1}(t), \dot{x}_{2}(t))=(u_{1}(t), u_{2}(t))$, where $x=(x_{1}, x_{2})\in\mathbb{R}^{2}$ is the system state being the position of the autonomous vehicle, and $u=(u_{1}, u_{2})\in\mathbb{R}^{2}$ is the control input being the velocities of the wheels. Let $x\in X:=[-15, 15]\times[-10, 10]$ and $u\in U:=[-2, 2]\times[-2, 2]$. Here, our objective is to design a controller such that the autonomous vehicle moves to a specific target region by avoiding four obstacles, where the target region is $\mathcal{O}_{\mathfrak{t}}=[12.7, 14.3]\times[7.2, 8.8]$, and the obstacles are denoted as the grey regions in Fig. \ref{fig-5}. Therefore, the specification is formalized as the LTL formula: $\phi=\lozenge\mathcal{O}_{\mathfrak{t}}\wedge\square\neg(\mathcal{O}_{1}\vee\mathcal{O}_{2}\vee\mathcal{O}_{3}\vee\mathcal{O}_{4})$. In this example, the vehicle is initialized in $X_{0}=[-13.3, -12.1]\times[-7.8, -6.6]$.

We first apply Algorithms \ref{alg-1}-\ref{alg-2} to partition the state space, which is shown in Fig. \ref{fig-4}. The partition results in $N=4$ zonotopes and $M=13$ constrained zonotopes. Based on intersection relation and Algorithm \ref{alg-3}, these zonotopes and constrained zonotopes are labeled into 17 symbols (i.e., $\Pi=\{\pi_{i}: i=1, \ldots, 17\}$), and a graph $\mathcal{G}=(\mathcal{V}, \mathcal{E})$ is generated (see Fig. \ref{fig-4}). The graph $\mathcal{G}$ is generalized by including the initial state set $X_{0}$ and the target set $\mathcal{O}_{\mathfrak{t}}$, and further a path $\mathbf{P}$ can be found via graph techniques. In Fig. \ref{fig-4}, both red and black lines are paths, and we choose the red one including 2 zonotopes and 2 constrained zonotopes, which are denoted as an ordered sequence $\mathbf{Z}_{\textsf{p}}:=\{\mathbf{Z}^{\cstr}_{1}, \mathbf{Z}_{1}, \mathbf{Z}_{2}, \mathbf{Z}^{\cstr}_{2}\}$.

\begin{table}[tp]
\centering
\caption[caption]{\scriptsize{Comparison of transition numbers and run times}\footnotemark[1]}
\label{tab-1}
\begin{center}
\begin{tabular}{c|c|c}
\hline
& Classic approach\footnotemark[2]  & Our approach \\ \hline \xrowht{4pt}
transition number  & 662238 & 383252 \\ \hline \xrowht{4pt}
$\mathfrak{t}_{\textrm{abs}}$  & 9.9094s  &2.0905s \\
 \hline \xrowht{4pt}
$\mathfrak{t}_{\textrm{con}}$ &0.2514s  &0.1520s \\ \hline
\end{tabular}
\end{center}
\end{table}
\footnotetext[1]{\scriptsize{$\mathfrak{t}_{\textrm{abs}}$ and $\mathfrak{t}_{\textrm{con}}$ denote the computation times of the abstraction construction and control synthesis, respectively.}}
\footnotetext[2]{\scriptsize{The classic approach is similar to the one in \cite{Reissig2017feedback} and the related link is here: https://github.com/dionysos-dev/Dionysos.jl.}}

The second step is the controller design. Let $\tau=0.5$s, and the set $U$ is approximated as $U_{2}:=\{0.5i: i\in\{-4, -3, \ldots, 3, 4\}\}^{2}$. Different (constrained) zonotopes can be approximated with different state space parameters, which are constrained by the desired precision $\varepsilon=1$. The state space parameters are chosen as $0.4, 0.8, 0.5, 0.4$ with the corresponding (constrained) zonotopes. Hence, the symbolic abstraction is constructed for each (constrained) zonotope, and the abstract controller is designed. For instance, in $\mathbf{Z}^{\cstr}_{2}$, the target region is $\mathbf{Z}^{\cstr}_{2}\cap\mathcal{O}_{\mathfrak{t}}$, and the initial region is $\mathbf{Z}^{\cstr}_{2}\cap\mathbf{Z}_{2}$. It takes 0.1563s to construct the abstraction with $36729$ transitions and 0.004425s to derive the abstract controller. The state trajectory is shown in Fig. \ref{fig-5}.

Comparing with the existing works \cite{Girard2012controller, Ren2020symbolic, Pola2008approximately, Reissig2017feedback} where the whole state space is approximated, the partition-based specification verification results in finite (constrained) zonotopes, which are approximated only in this paper. In terms of the transition number and computation times, the comparison between our approach and the bottom-up approach (e.g., \cite{Reissig2017feedback}) is shown in Table \ref{tab-1}.

\section{Conclusion}
\label{sec-conclusion}

In this paper, we addressed the control design problem for nonlinear control systems with temporal logic specifications. We first proposed a novel method for the partition of the state space to verify whether the desired specification can be satisfied in the state space, and then applied abstraction-based techniques to explore the control design based on the specification verification. Finally, we presented a numerical example to demonstrate the proposed control strategy. Future work will study the application of the proposed approach to complex specifications like spatio-temporal logic.

%%%%%%%%%%%%%%%%%%%%%%%%%%%%%%%%%%%%%%%%%%%%%%%%%%%%%%%%%%%%%%%%%%%%%%%%%%%%%%%%%%%%%%%%%%%%%%%%%%%%%%%%%%%%%%%%%


\end{document}